\documentclass[aps,pre,showpacs,twocolumn,"Cornell University, Ithaca, NY, 14853"]{revtex4}

\usepackage{amsmath}
\usepackage{empheq}
\usepackage[parfill]{parskip}
\usepackage{graphicx}
\usepackage[justification=raggedright,singlelinecheck=false]{caption}
\usepackage{color}
\usepackage{soul}

\begin{document}
\newcommand{\bi}{\begin{itemize}}
\newcommand{\ei}{\end{itemize}}
\newcommand{\be}{\begin{equation}}
\newcommand{\fe}{\end{equation}}
\newcommand{\noi}{\noindent}

\setlength\parindent{20pt}

\title{Dynamics of a population of oscillatory and excitable elements}

\author{Kevin P.  O'Keeffe and Steven H. Strogatz}
\affiliation{Center for Applied Mathematics, Cornell University, Ithaca, NY 14853, USA}

\date{\today}

\begin{abstract}
We analyze a variant of a model proposed by Kuramoto, Shinomoto, and Sakaguchi for a large population of coupled oscillatory and excitable elements. Using the Ott-Antonsen ansatz, we reduce the behavior of the population to a two-dimensional dynamical system with three parameters.  We present the stability diagram and calculate several of its bifurcation curves analytically, for both excitatory and inhibitory coupling. Our main result is that when the coupling function is broad, the system can display bistability between steady states of constant high and low activity, whereas when the coupling function is narrow and inhibitory, one of the  states in the bistable regime can show persistent pulsations in activity. 

\end{abstract}

\maketitle

\section{Introduction}

In 2008, Ott and Antonsen~\cite{OA} discovered a remarkable ansatz that reduces the infinite-dimensional dynamics of the Kuramoto model of coupled oscillators~\cite{kuramoto} to a flow on a phase plane.  Their ansatz has since been used to shed light on diverse physical and biological systems \cite{pikovsky2015}, ranging from pedestrian-induced oscillations of wobbly footbridges \cite{abdulrehem2009} to arrays of Josephson junctions \cite{marvel2009} and periodically forced circadian rhythms \cite{childs2008}. 

More recently, several authors have shown how to use the Ott-Antonsen ansatz to derive exact firing rate equations for a large population of spiking neurons~\cite{pazo1, pazo2, so_barreto1, Laing_neural_field_model}. The approach relies on approximating the neurons as oscillatory or excitable elements described by a single phase variable. An early forerunner of this idea was proposed by Kuramoto, Shinomoto, and Sakaguchi \cite{kuramoto1}. In 1987, they considered a radically simplified model in which each neuron was modeled as ``the simplest possible excitable system'' \cite{kuramoto1}, coupled by delta function pulses.  Their governing equation is 
\begin{equation}
\dot \phi_i = \omega_i- b\sin \phi_i + \frac{K}{N}\sum_{j=1}^N \delta(\phi_j + \pi/2)
\label{kuramoto_original}
\end{equation}
for $i=1, \ldots, N$.  The natural frequencies $\omega_i$ are assumed to be uniformly distributed on the interval $0 < \omega_i < b$ so that the individual elements are excitable rather than spontaneously oscillatory, $K>0$ is the  coupling strength, and $\phi = -\pi/2$ is the phase at which the model neuron fires.  By using a self-consistency argument to find the stationary states in the limit $N \rightarrow \infty$, Kuramoto, Shinomoto, and Sakaguchi \cite{kuramoto1} showed the system could be bistable: either all the neurons could be off (not firing) or most could be on (firing repetitively), depending on the initial conditions. Their self-consistency analysis also predicted the collective firing rate.  However, given the tools available at the time, they could not analyze the model's dynamics, stability, or bifurcations.  

We were curious to revisit this problem, armed with the Ott-Antonsen ansatz. Rather than aim for biological realism, we study a model close in structure to Eq.~\eqref{kuramoto_original}. Our motivation is theoretical, namely, to explore the model as a dynamical system.


\section{The model} 

\begin{figure}[h]
\centerline{\includegraphics[width=3.8cm]{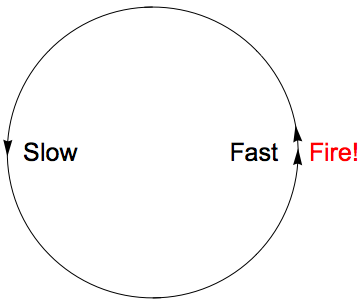}} 
\caption{\label{schematic} Schematic of each oscillator's behavior on the unit circle. Oscillators move slowly on the circle's left hand side and rapidly on its right hand side. Oscillators also fire a pulse whenever they pass through $\theta = 0$. }
\end{figure}

The model we study is
\be
\dot{\theta_i} = \omega_i + b \cos \theta_i + \frac{K}{N} \sum_{j=1}^N P(\theta_j)
\label{EOM}
\fe 
for $i = 1, \ldots, N$, where $N \gg 1$. Here $\theta_i$ and $\omega_i$ are the phase and natural frequency of oscillator $i$ and $K$ is the coupling strength. (Note that for convenience we have defined the phase $\theta_i = \phi_i+\pi/2$ relative to the notation in Eq.~\eqref{kuramoto_original}, so that the firing phase now corresponds to $\theta = 0$, and $-b \sin \phi_i$ transforms into $b \cos \theta_i$.)   The term $b \cos \theta$ introduces nonlinearities into each oscillator's intrinsic cycle. It slows the oscillators down on $(\pi/2, 3\pi /2)$, and speeds them up on $(-\pi/2, \pi/2 )$. This behavior is shown schematically in Fig.~\ref{schematic}. 

The inclusion of the $b \cos \theta$ term splits the population into two types of oscillators:  \textit{excitable} and \textit{self-oscillatory}. Such systems have been previously considered by Daido, Pazo, Montbrio and others \cite{pazo3, daido_ageing, daido_excitable}.  The excitable oscillators are those with $|\omega_i| < b$. In the absence of coupling ($K=0$), each oscillator has a stable equilibrium state on its circle, corresponding to the resting state of a neuron, as well as an unstable state,  corresponding to the firing threshold. An excitable oscillator perturbed past its firing threshold will go on a long excursion around its state circle, akin to a neuron being excited to fire an action potential before returning to rest. The self-oscillatory elements are those with $|\omega_i| > b$; when $K=0$ they oscillate spontaneously, emitting pulses periodically whenever they pass through $\theta=0$ on the circle. 

Following Ott and Antonsen~\cite{OA}, we assume the $\omega_i$ are drawn from a Lorentzian distribution with center frequency $\mu$ and width $\gamma$, given by the density 
\be
g(\omega) = \frac{\gamma}{\pi} \frac{1}{(\omega - \mu)^2 + \gamma^2}. 
\label{Lorentzian}
\fe
Although this assumption differs from the uniform distribution assumed by Kuramoto, Shinomoto, and Sakaguchi \cite{kuramoto1}, it has the advantage of greater analytical tractability. 

The coupling in Eq.~\eqref{EOM} is mediated through an influence function $P(\theta_j)$, assumed to   be a unimodal, symmetric, nonincreasing function centered at $\theta=0$. We analyze two extreme cases: a broad function $P(\theta) = 1+ \cos \theta$, and a narrow function $P(\theta) = \delta(\theta)$. Intuitively, $P(\theta)$ represents how one oscillator's activity affects all the others. When $P(\theta) = 1+ \cos \theta$, an oscillator's influence waxes and wanes gradually over its cycle, achieving a maximum at $\theta = 0$. But when $P(\theta) = \delta(\theta)$, the influence is sudden; it occurs precisely when an oscillator crosses $\theta = 0$, at which time it fires a sharp pulse.  Depending on the sign of $K$, this pulse can be either excitatory ($K>0$) or inhibitory ($K<0$). 

When the system is uncoupled ($K=0$), its long-term behavior is clear: the excitable elements remain at their stable rest states while the self-oscillators fire periodically but ineffectually.  When $K \ne 0$, however, the excitable elements feel the pulses of the self-oscillators, and the collective dynamics are no longer as clear. Will the excitable oscillators remain stuck at rest, or start firing periodically themselves? Or perhaps more complicated behavior will arise. Our goal is to answer these questions.


\section{Results}

\subsection{Numerical results}
\begin{figure}[h]
\centerline{\includegraphics[width=4.0cm]{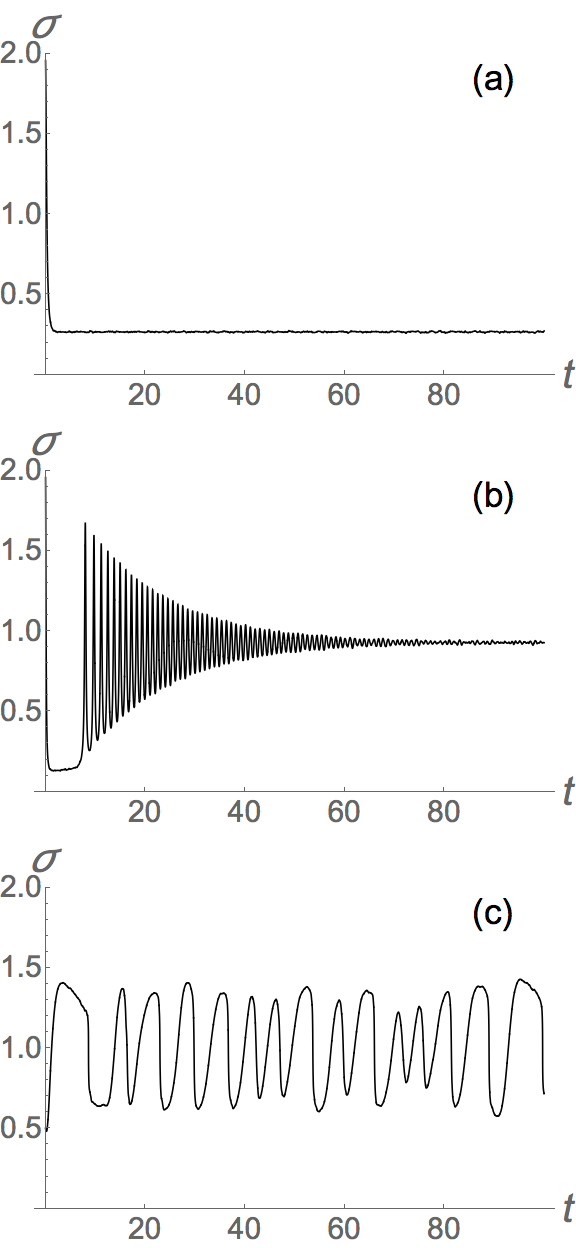}} 
\caption{\label{activity_time_series} 
Activity time series in the three forms of long-term behavior for Eq.~\eqref{EOM}. (a) Low activity: $(K, \gamma, \mu) = (3,0.05,0)$ with $P(\theta) = 1 + \cos \theta$. (b) High activity: $(K, \gamma, \mu) = (7.5, 0.05, 0)$ with $P(\theta) = 1 + \cos \theta$. (c) Oscillatory activity: $(K, \gamma, \mu) = (-4.15,0.01,0)$ with $P(\theta) = \delta(\theta)$. All simulations were made with $N = 5000$ oscillators using a fourth-order Runge-Kutta method with a timestep of $0.01$. }
\end{figure}

Numerical integration of Eq.~\eqref{EOM} indicates that the system displays three kinds of long-term behavior, depending on the choice of parameters. We characterize these states by their macroscopic \emph{activity},  which following Ref.~\cite{kuramoto1} we define as 
\be
\sigma(t) =  \frac{1}{N}\sum_{j=1}^N P(\theta_j(t)). 
\label{activity}
\fe
The activity is simply the average of the instantaneous pulse strength, and can be thought of as a current which drives the oscillators. The three states may be characterized as low activity [Fig.~\ref{activity_time_series}(a)], high activity [Fig.~\ref{activity_time_series}(b)], and oscillatory activity [Fig.~\ref{activity_time_series}(c)]. 

\begin{figure}[h!]
\centerline{\includegraphics[width=9.0cm]{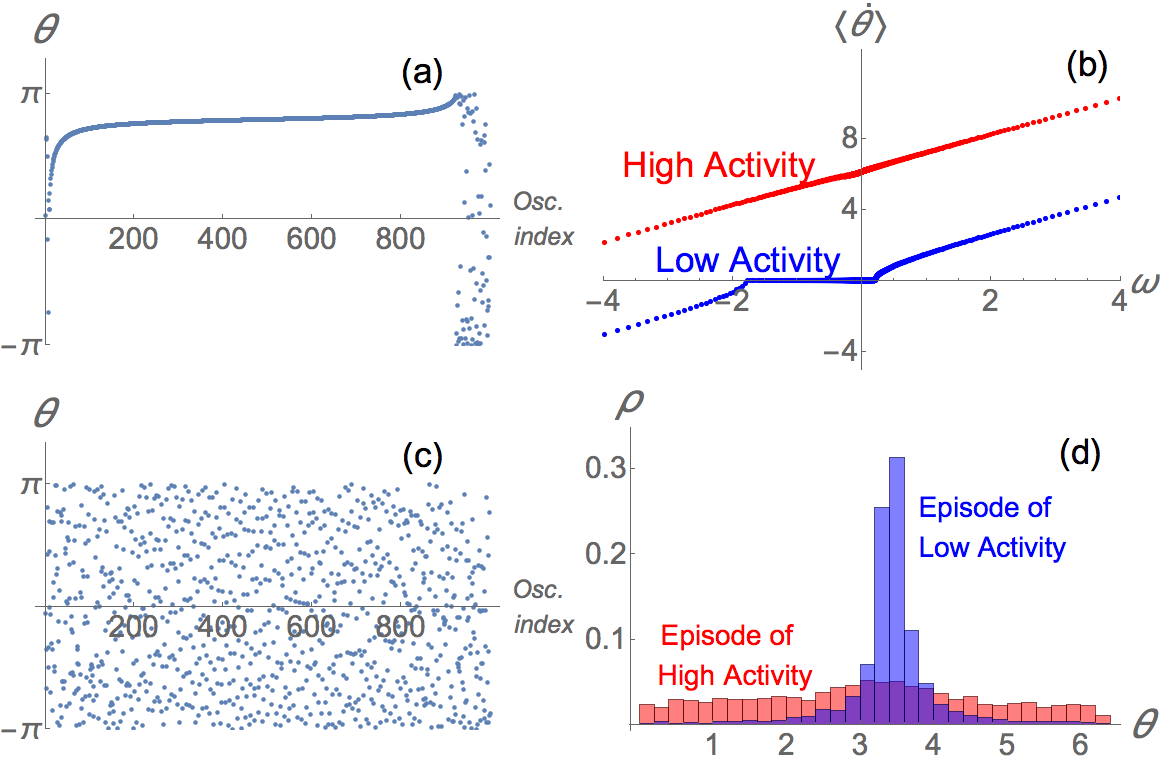}} 
\caption{\label{numerical_states} Snapshots of phase space distributions and average frequencies for three states. (a), (b): Low activity: the phases of the oscillators in the middle of the distribution are at rest, while those in the tails rotate periodically around the unit circle. Parameter values: $(K, \gamma, \mu) = (3.5, 0.05,0)$ with $P(\theta) = 1 + \cos \theta$. (b), (c): High activity: All oscillators rotate periodically around the unit circle, forming a scattered phase distribution. Parameter values: $(K, \gamma, \mu) = (7.5, 0.2, 0)$ with $P(\theta) = 1 + \cos \theta$. (d) Oscillatory activity: The oscillators alternate between episodes of high and low activity. In the low activity episode, the oscillators pile up and form a peaked distribution on $\pi/2 < \theta < 3\pi/2$. Then they quickly pass through $3 \pi/2 < \theta < 5\pi/2$ creating an episode of high activity. Parameter values:  $(K, \gamma, \mu) = (-4.15,0.01,0)$ with $P(\theta) = \delta(\theta)$. All data were obtained by integrating Eq.~\eqref{EOM} for $N = 1000$ oscillators using a fourth-order Runge-Kutta method with a timestep of $0.01$. }
\end{figure}

It is instructive to consider the corresponding behavior in state space. Figure~\ref{numerical_states}(a) shows a snapshot of the phases of the oscillators in the low activity state. The oscillators in the middle of the frequency distribution are stationary and never fire. Those in the tails, however, execute full cycles, which is why their phases appear scattered. This periodic behavior is also evident in Figure~\ref{numerical_states}(b), which plots the average frequency of oscillators $\langle \dot{\theta} \rangle$ in the low activity state versus their natural frequency $\omega$. Note the oscillators in the middle of the distribution form a plateau at $\langle \dot{\theta} \rangle = 0$, meaning that they are not firing, while the oscillators in the tails have have $\langle \dot{\theta} \rangle \ne 0$. This low activity state is achieved for both the broad influence function $P(\theta) = 1 + \cos \theta$ and the narrow pulse $P(\theta) = \delta(\theta)$.

For the high activity state, Figs.~\ref{numerical_states}(b) and \ref{numerical_states}(c) show that most oscillators are running around the circle, firing repetitively, leading to a nonzero average frequency for them. This high activity state is also achieved for both the broad and narrow influence functions.
 
In the oscillatory activity state, the oscillators perform complicated movements, leading to roughly periodic fluctuations in $\sigma(t)$. In particular, there is a slow-fast structure. The oscillators first go through a low activity phase as they slowly move through the states $\pi/2 < \theta < 3\pi/2$ on the left hand side of the state circle, creating a peaked density there. They then quickly pass through $3\pi/2 < \theta < 5\pi/2$ (the right hand side of the unit circle), creating an episode of high activity. Figure~\ref{numerical_states}(d) shows the density $\rho(\theta)$ of oscillators during these episodes of low and high activity. This oscillatory activity state is achieved only for $P(\theta) = \delta(\theta)$.

\subsection{Reduction to low-dimensional system}
We turn now to the analysis. In the $N \rightarrow \infty$ limit, the system \eqref{EOM} can be analysed with the  Ott-Antonsen ansatz. Since applying this ansatz has become standard, we give an abridged derivation here, and direct the reader to \cite{OA} for a fuller account.

In the infinite-$N$ limit, we describe our system by a density $\rho(\theta, \omega, t)$, where $\rho(\theta, \omega, t) d\theta$ gives the fraction of oscillators with natural frequency $\omega$ that have phases between $\theta$ and $\theta + d \theta$ at time $t$. The Ott-Antonsen ansatz is then 
\be
\rho(\theta, \omega, t) = \frac{1}{2\pi} \left[1  + \sum_{n=1}^{\infty} \bar{\alpha}(\omega,t)^n e^{ i n \theta} + c.c. \right]  
\label{oa_ansatz}
\fe

\noindent
where the overbar notation and c.c. both denote complex conjugation. The main result of  Ott-Antonsen theory is that densities of the above form constitute an invariant manifold that determines the system's long-term dynamics. We therefore restrict our attention to this special manifold, where, as we will show, the system is easily analyzed.

The density \eqref{oa_ansatz} obeys the continuity equation
\be
\frac{\partial \rho}{ \partial t} + \frac{\partial}{\partial \theta}( v \rho  ) = 0 
\label{cont_eqn}
\fe
where the velocity $v = \omega + b \cos\theta + K \sigma$ is given by the right hand side of \eqref{EOM},  and is interpreted in the Eulerian sense. Substituting the Ott-Antonsen ansatz \eqref{oa_ansatz} into the continuity equation \eqref{cont_eqn} yields the following ordinary differential equation (ODE) for $\alpha(\omega, t)$:
\be
\dot{\alpha} =  i \omega \alpha + i \alpha K \sigma(\alpha) + \frac{1}{2} i b (\alpha^2 + 1).
\label{alpha_eqn1}
\fe

\noindent
This is an infinite-dimensional set of ODEs, one for each natural frequency $\omega$. 

The macroscopic behavior of the system, however, has much lower dimensional dynamics. This macroscopic behavior can be described using the complex Kuramoto order parameter
\be
Z = \langle e^{i \theta} \rangle = \int e^{i \theta}  \rho(\theta, \omega, t) g(\omega) d \theta d \omega.
\label{Z}
\fe

\noindent 
Inserting the  Ott-Antonsen ansatz \eqref{oa_ansatz} into this integral, and doing the integration over $\theta$, leads to $Z = \int \alpha(\omega, t) g(\omega) d \omega$. This integral can in turn be calculated by extending $\omega$ into the complex plane, and computing a contour integral over an infinitely large semicircle in the upper half plane. By assuming that $\alpha$ has the required analytic properties, and by noting that the Lorentzian distribution \eqref{Lorentzian} has a simple pole at $\omega = \mu + i \gamma$, we can use the residue theorem to compute the integral for $Z$. This leads to
\be
Z = \alpha(\omega = \mu + i \gamma, t ).
\label{alpha_eqn2}
\fe

\noindent
Hence the order parameter is simply $\alpha$ evaluated at the complex frequency $\omega = \mu + i \gamma$. This remarkable fact, in concert with \eqref{alpha_eqn1}, lets us analyze the behavior of $Z$. Setting $Z = r e^{i \phi}$, evaluating \eqref{alpha_eqn1} at $\omega = \mu + i \gamma$, and collecting real and imaginary parts, yields the following ODEs for $r$ and  $\phi$:

\begin{align}
\dot{r} &= - \gamma r + \frac{b}{2} ( 1 - r^2) \sin \phi  \nonumber \\
\dot{\phi} &= \mu + K \sigma(r, \phi) + \frac{b}{2} (r + r^{-1}) \cos \phi. 
\label{odes}
\end{align}

\noindent
Note that the equations \eqref{odes} hold for a general influence function $P(\theta)$, whose presence is implicit in the activity $\sigma(r,\phi)$. We next analyze these ODEs for specific instances of $P(\theta)$.


\subsection{Broad Pulse}

Consider the case $P(\theta) = 1 + \cos \theta$. The activity is given by $\sigma = N^{-1} \sum_j (1 + \cos \theta_j )= 1 + \text{Re} \left( N^{-1} \sum_j e^{i \theta_j} \right)$. Remembering $Z = \langle e^{i \theta} \rangle$, we get $\sigma = 1 + \text{Re}(Z)$, or
\be
\sigma_{broad} = 1 + r \cos \phi.
\label{sigma_broad}
\fe

\noindent
We next nondimensionalize the system. By rescaling time, we set $b = 1$ without loss of generality (so that the remaining parameters $K,\mu,\gamma$ are measured in units of $b$). Then, substituting \eqref{sigma_broad} into \eqref{odes} results in the following set of ODEs for $r$ and $\phi$:  
\begin{align}
\dot{r} &= - \gamma r + \frac{1}{2} ( 1 - r^2) \sin \phi  \nonumber \\
\dot{\phi} &= \mu + K(1 + r \cos \phi) + \frac{1}{2} (r + r^{-1}) \cos \phi  .
\label{broad_pulse}
\end{align}

\noindent
The system \eqref{broad_pulse} has three parameters: the coupling strength $K$, the center frequency $\mu$ of the Lorentzian distribution, and its width $\gamma$. 

\begin{figure}[h!]
\centerline{\includegraphics[width=7.5cm]{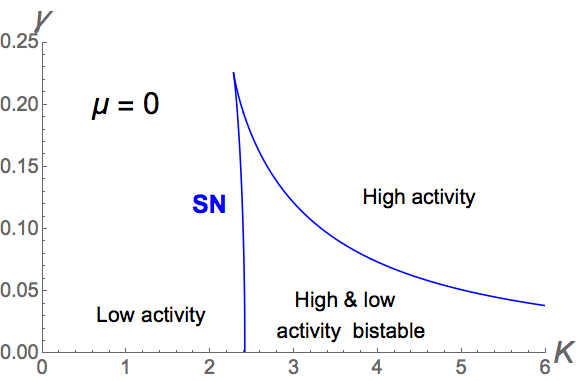}} 
\caption{\label{bif_diagram_broad_pulse} (Color online) Stability diagram for system \eqref{broad_pulse} when $\mu = 0$. The abbreviation SN stands for saddle-node bifurcation.  }
\end{figure}

We first set $\mu = 0$ to make the analysis as simple as possible. We found saddle-node curves by solving for the fixed points of \eqref{broad_pulse} and $\det(J)=0$ simultaneously using Mathematica. While the resulting expressions are analytic, they are rather cumbersome, so we omit showing them. These saddle-node curves are shown in Fig.~\ref{bif_diagram_broad_pulse}, and join at a cusp at  $(K, \gamma) \approx (2.27, 0.22)$. They thus define a parameter region in which both the high and low activity states are locally stable. 

This region of bistability is the counterpart of that  anticipated by Kuramoto, Shinomoto, and Sakaguchi~\cite{kuramoto1}. In their model~\eqref{kuramoto_original}, they were able to calculate the self-consistent levels of steady-state activity, but now, with the benefit of the Ott-Antonsen approach, we can prove the stability of those states and derive the boundaries of the bistable region exactly.

\subsubsection{Stability diagram for $\mu > 0$}
\begin{figure}[h]
\centerline{\includegraphics[width=8cm]{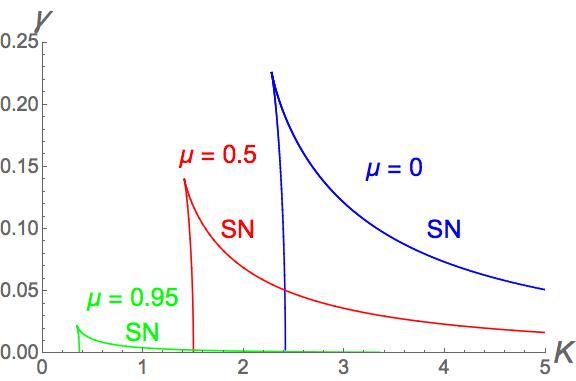}} 
\caption{\label{shrinking_of_bistable} (Color online) Illustration of the shrinking of the bistable region defined by the saddle-node (SN) curves as $\mu$ is increased from $0$. The region fully disappears at $(K, \gamma, \mu) = (0,0,1)$. }
\end{figure}

The Ott-Antonsen approach also lets us explore phenomena in parameter regions beyond the scope of the methods used in Ref.~\cite{kuramoto1}. We first increase $\mu$ from $0$. As shown in Fig.~\ref{shrinking_of_bistable}, the bistable region shrinks until it disappears at $(K, \gamma, \mu_c) = (0,0,1)$ , a result we obtained numerically. In the original units, $\mu_{c_+} = b$, indicating that bistability disappears when $ \mu = \langle \omega_i \rangle \geq b$, or in other words, when the average oscillator is of the spontaneously firing variety.

\begin{figure}[h!]
\centerline{\includegraphics[width=7.5cm]{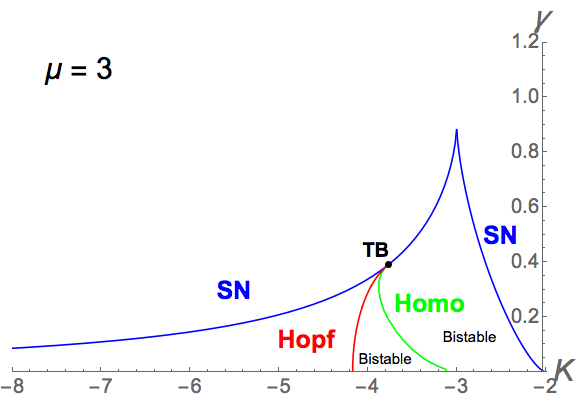}} 
\caption{\label{bif_diagram_broad_pulse_mu_3} (Color online)  Stability diagram for system \eqref{broad_pulse} when $\mu =3$. The abbreviations SN, Homo, and TB stand for saddle-node, homoclinic, and Takens-Bogdanov bifurcations, respectively. The Hopf curve is subcritical.}
\end{figure}

If we increase $\mu$ past $\mu_c = 1$, we get a rich sequence of bifurcations, but now for $K < 0$, corresponding to inhibitory coupling.  As shown in Fig.~\ref{bif_diagram_broad_pulse_mu_3}, another pair of saddle-node bifurcation curves meet in a cusp catastrophe. There is also a curve of subcritical Hopf bifurcations, which meet the saddle-node curves at a Takens-Bogdanov point. These features were all obtained analytically, but the resulting expressions are too complicated to be presented here, except for the Hopf curve, which is given by
\be
\gamma = \frac{(K+2) \sqrt{(4 K+5) K^2+4 (K+1) \mu ^2+8 (K+1) K \mu }}{2 K \sqrt{-K-1}}.
\fe

\noindent
The presence of a Takens-Bogdanov bifurcation implies the existence of a curve of homoclinic bifurcations, which we computed numerically.

\begin{figure}[h!]
\centerline{\includegraphics[width=8.5cm]{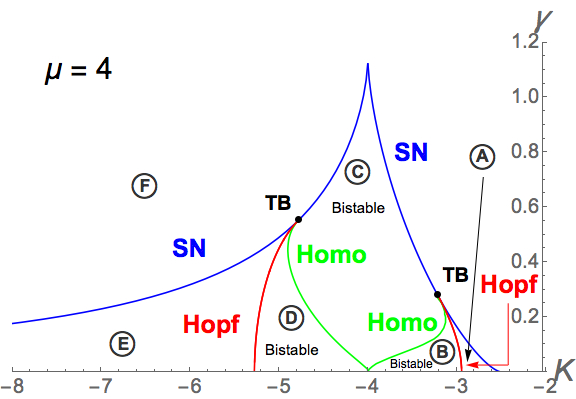}} 
\caption{\label{bif_diagram_broad_pulse_mu_4} (Color online)  Stability diagram for system \eqref{broad_pulse} when $\mu =4$. The abbreviations SN, Homo, and TB stand for saddle-node, homoclinic, and Takens-Bogdanov bifurcations, respectively. The Hopf curves are subcritical. }
\end{figure}

\begin{figure}[h!]
\centerline{\includegraphics[width=7.25cm]{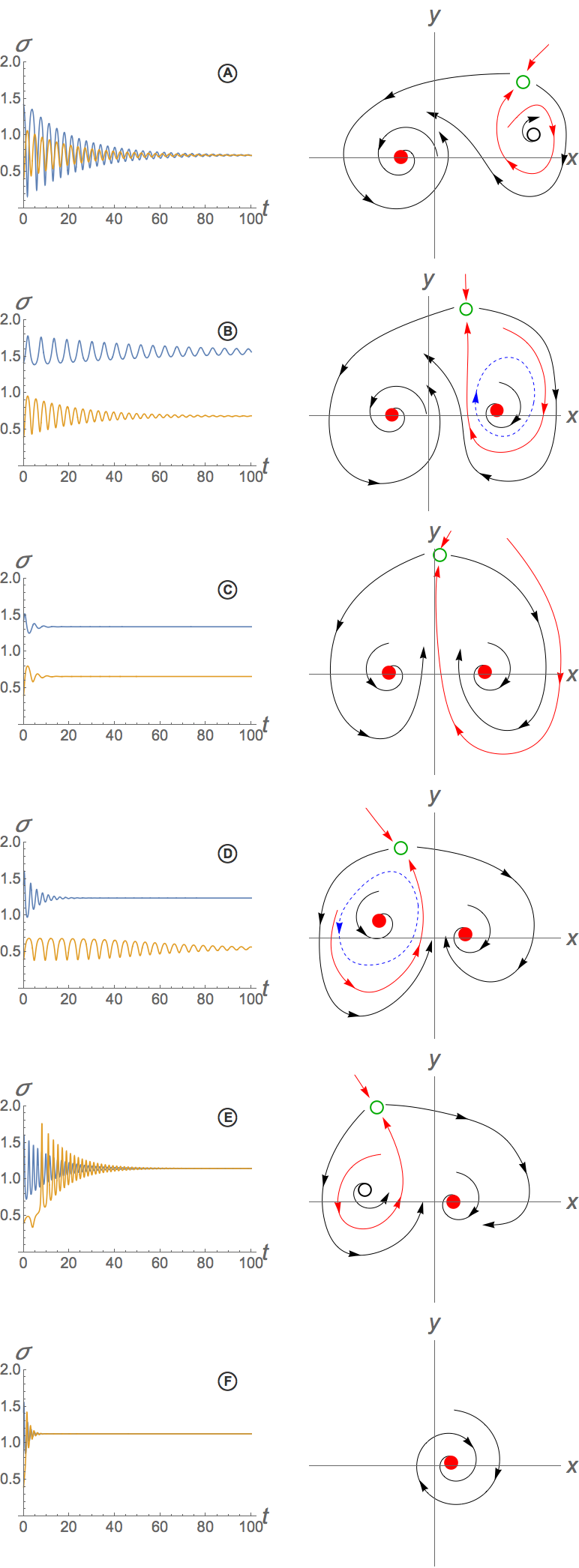}} 
\caption{\label{activity_phase_plane} (Color online)  Left: Activity time series for regions A through F  depicted in Fig.~\ref{bif_diagram_broad_pulse_mu_4}. Right: Schematic phase portraits for each of these regions. Red, white, and green points denote stable, unstable, and saddle fixed points. Unstable limit cycles are dashed and plotted blue. Stable manifolds of saddle points are plotted red.}
\end{figure}

Recall that when $0 \leq \mu < 1$ (Figs.~\ref{bif_diagram_broad_pulse} and \ref{shrinking_of_bistable}), the saddle-node curves define a region of bistability between the high and low activity states. The same bistability holds when $\mu > 1$. However, the Hopf and homoclinic bifurcation curves complicate the picture by creating smaller subregions of bistability, as shown in Fig. \ref{bif_diagram_broad_pulse_mu_3}. As we increase $\mu$ further, \textit{another} Takens-Bogdanov point appears, along with its required homoclinic and Hopf bifurcation curves. This scenario is shown in Fig. \ref{bif_diagram_broad_pulse_mu_4}. The medley of bifurcation curves define six distinct regions in the $(K, \gamma)$ plane. In Fig. \ref{activity_phase_plane} we show the phase portraits in each of these regions. Also shown are the time series for the activity $\sigma(t)$ as per \eqref{sigma_broad}. In every case, the system is either monostable or bistable.

\subsubsection{ Stability diagram for $\mu < 0$}

\begin{figure}[h!]
\centerline{\includegraphics[width=7.5cm]{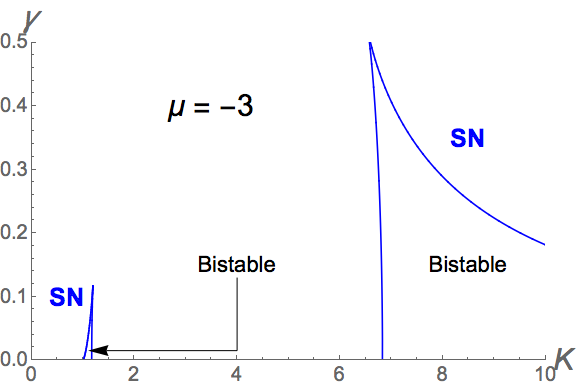}} 
\caption{\label{bif_diagram_broad_pulse_big_mu} (Color online)  Stability diagram for system \eqref{broad_pulse} when $\mu = -3$. The abbreviation SN stands for saddle-node. As can be seen, there are two regions of bistability.}
\end{figure}

Will the system bifurcate in such complicated ways for $\mu < 0$? For $-1 \leq \mu \leq 0$, the picture is the same as in Fig.~\ref{bif_diagram_broad_pulse}. Just as increasing $\mu$ made the bistable region smaller, decreasing $\mu$ makes it bigger. But in contrast to $\mu > 0 $, there are no Hopf curves, and no Takens-Bogdanov points associated with them. However for $\mu < -1$, a second bistable region is born (Fig.~\ref{bif_diagram_broad_pulse_big_mu}).

\subsubsection{Summary}

\begin{figure}[h!]
\centerline{\includegraphics[width=8.5cm]{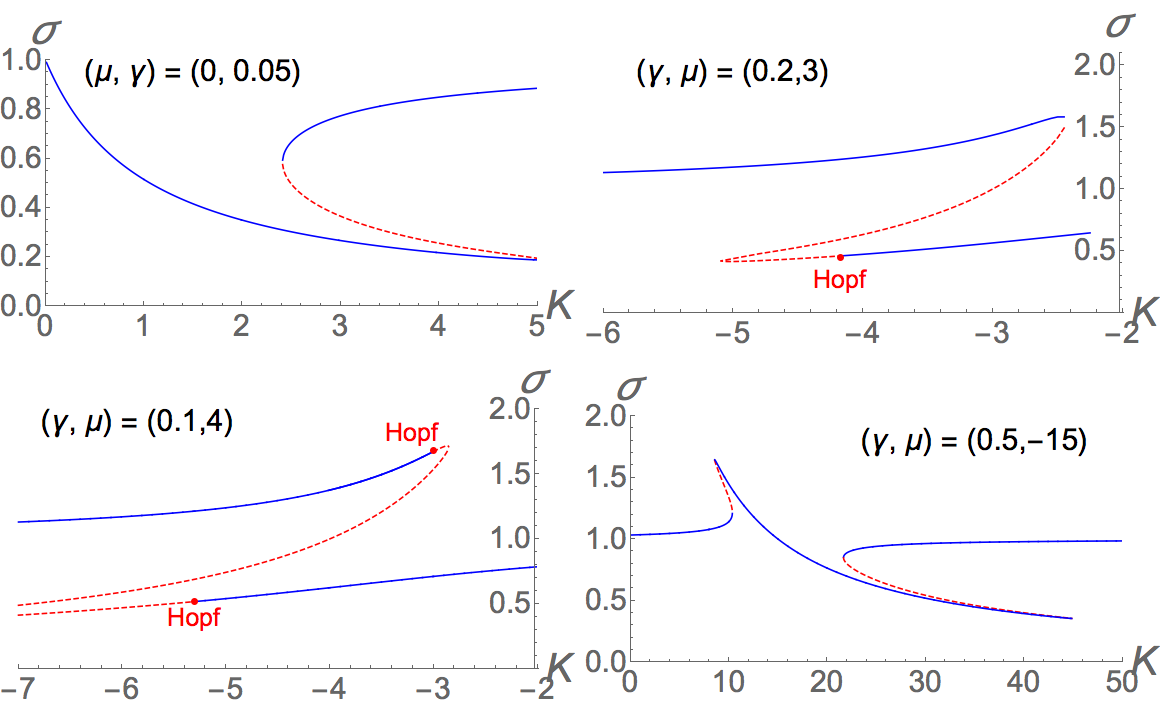}} 
\caption{\label{activity_broad_pulse} (Color online) Activity $\sigma(K)$ for various $(\mu, \gamma)$. Blue, solid curves denote stable branches, and red, dashed curves denote unstable branches. The top left and right panels correspond to Figures  \ref{bif_diagram_broad_pulse} and \ref{bif_diagram_broad_pulse_mu_3}, while the bottom left and right panels correspond to Figures \ref{bif_diagram_broad_pulse_mu_4} and \ref{bif_diagram_broad_pulse_big_mu}.
}
\end{figure}

Let us distill the results so far. A three-parameter bifurcation study of the system \eqref{broad_pulse} yields stability diagrams in the $(K, \gamma)$ plane for different values of $\mu$. Although there are four dynamically distinct slices of the $(K,\gamma)$ plane, as shown in  Figs. \ref{bif_diagram_broad_pulse}, \ref{bif_diagram_broad_pulse_mu_3}, \ref{bif_diagram_broad_pulse_mu_4}, and \ref{bif_diagram_broad_pulse_big_mu}, they all tell the same story:  with the broad coupling used in \eqref{broad_pulse}, the system \eqref{EOM} always reaches a steady state of constant activity $\sigma$. For some parameter values, the system is bistable, with high and low activity states coexisting.  Figure \ref{activity_broad_pulse} summarizes how the steady-state level of activity $\sigma$ depends on the coupling $K$ for the four dynamically distinct slices of $(K, \gamma)$ space.

\subsection{Narrow Pulse}
How generic is the behavior of the broad pulse system \eqref{broad_pulse}? To answer this question, we investigate a different choice of the influence function $P(\theta)$. We change the broad pulse to an infinitesimally narrow pulse, $P(\theta) = \delta(\theta)$, so that oscillators fire  precisely when they are at $\theta = 0$. With this choice of influence function, our model \eqref{EOM} is very similar to that studied by Kuramoto, Shinomoto, and Sakaguchi \cite{kuramoto1}. The differences are that they restricted attention to excitatory coupling $K > 0$ and assumed a uniform distribution of frequencies on the interval $0 \leq \omega \leq b $. Using a self-consistency analysis, they established the existence of states with high and low activity, and identified parameters at which those states are bistable. Now, with the benefit of  Ott-Antonsen theory, the rest of the  bifurcation diagram can be filled in. This leads to the discovery of a nonstationary state, in which the activity $\sigma(t)$ oscillates persistently. 
 
To begin the analysis, recall that the activity is given by $\sigma(t) = N^{-1} \sum_j P(\theta_j(t))$. In the infinite-$N$ limit, this becomes $ \sigma(t)= \int \delta(\theta) g(\omega) \rho(\theta, \omega, t) d \omega d \theta$.  The integral over $\theta$ is trivial, while that over $\omega$ can be computed by the residue theorem, giving $\sigma(t) = \rho(\theta = 0, \omega = \mu + i \gamma, t)$. Thus the activity is simply the time-dependent density of oscillators at $\theta = 0$.

An expression for $\sigma(t) = \rho(0, \mu + i \gamma, t)$ can be obtained as follows. When we introduced the  Ott-Antonsen ansatz \eqref{oa_ansatz} for $\rho$, we expressed it as a Fourier series. Summing this series gives $\rho(\theta, \omega, t) = (2 \pi)^{-1}\left[  (1 - |\alpha|^2) / ( 1 + 2 r \cos (\arg(\alpha) - \theta)) + |\alpha|^2) \right]$. Setting $\theta = 0$, $\omega = \mu + i \gamma$, and remembering $Z = r e^{i \phi} = \alpha (\omega + i \gamma)$ yields
\be
\sigma_{narrow} = \frac{1}{2 \pi } \frac{1 - r^2}{ 1 - 2 r \cos \phi  + r^2}.
\label{sigma_narrow}
\fe

\noindent
Plugging this into \eqref{odes} yields a two-dimensional dynamical system:
\begin{align}
\dot{r} &= - \gamma r + \frac{1}{2} ( 1 - r^2) \sin \phi \nonumber \\
\dot{\phi} &= \mu + \frac{K}{2 \pi} \frac{1 - r^2}{1 - 2 r \cos\phi + r^2}+ \frac{1}{2} (r + r^{-1}) \cos \phi.
\label{narrow_pulse}
\end{align}

\begin{figure}[h!]
\centerline{\includegraphics[width=9cm]{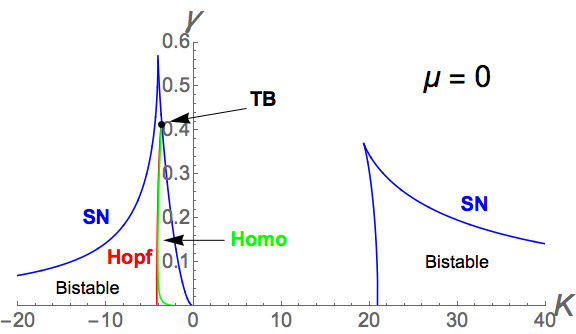}} 
\caption{\label{bif_diagram_narrow_pulse} 
Stability diagram for system \eqref{narrow_pulse} when $\mu = 0$. The abbreviations SN and TB stand for saddle node and Takens-Bogdanov bifurcations, respectively. The Hopf curve is supercritical. }
\end{figure}

\begin{figure}[h!]
\centerline{\includegraphics[width=9cm]{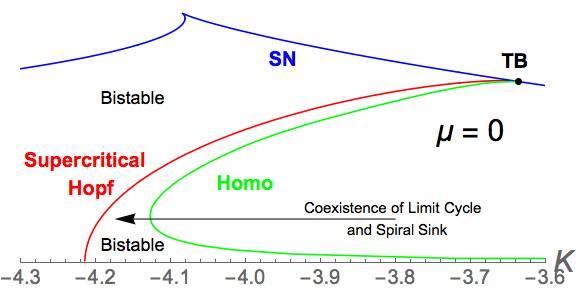}} 
\caption{\label{bif_diagram_narrow_pulse_zoomed} 
Enlarged portion of Figure \ref{bif_diagram_narrow_pulse} showing the supercritical Hopf and homoclinic bifurcation curves. }
\end{figure}

\begin{figure}[h!]
\centerline{\includegraphics[width=8cm]{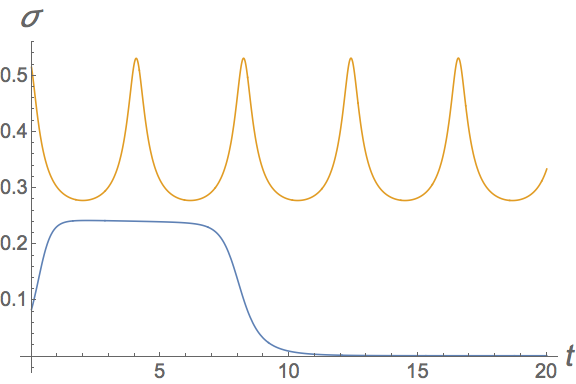}} 
\caption{\label{activity_narrow_pulse} 
Time series of activity defined by \eqref{sigma_narrow}, showing the coexistence of a stable limit cycle and a stationary state for $(K,\gamma, \mu) = (-4.15, 0.01, 0)$.}
\end{figure}

From here, we perform the same analysis as for the broad pulse system. The bifurcation diagram for $\mu = 0$ is shown in Fig.~\ref{bif_diagram_narrow_pulse}. It has the same features as those found in the broad pulse system (see Figs. \ref{bif_diagram_broad_pulse} and \ref{bif_diagram_broad_pulse_mu_3}), but with one important exception: the curve of Hopf bifurcations is now \emph{supercritical}, giving rise to a stable limit cycle. Figure~\ref{bif_diagram_narrow_pulse_zoomed} zooms in on the relevant part of Fig.~\ref{bif_diagram_narrow_pulse} to show the  homoclinic and Hopf curves more clearly. Note the qualitatively new kind of bistable region in which a stable limit cycle coexists with a stable spiral fixed point. Figure~\ref{activity_narrow_pulse} shows the behavior of $\sigma(t)$ in this region. The oscillatory activity state shown earlier in Fig.~\ref{activity_time_series}(c) corresponds to this limit cycle, but the simulation results of Fig.~\ref{activity_time_series}(c) do not display perfect periodicity because of finite-$N$ effects.

For nonzero $\mu$ we get the same behavior as for the broad pulse system: for increasing $\mu$, the bistability region again shrinks and finally disappears, similar to the scenario in Fig.~\ref{shrinking_of_bistable}. The creation of a series of Hopf curves and the resulting Takens-Bogdanov points also occurs, similar to Figs. \ref{bif_diagram_broad_pulse_mu_3} and \ref{bif_diagram_broad_pulse_mu_4}. However, in this case, the Hopf curves are supercritical. Lastly, for $\mu < 0$, two bistable regions occur in the $K>0$ plane, as in Fig.~ \ref{bif_diagram_broad_pulse_big_mu}.


\section{Conclusion}

Building on the work of Kuramoto, Shinomoto, and Sakaguchi \cite{kuramoto1}, we have studied a mean-field model of infinitely many excitable and self-oscillatory elements coupled by a pulsatile influence function. The Ott-Antonsen ansatz \cite{OA} allowed us to to reduce the system's macroscopic dynamics to two ordinary differential equations in three parameters $(K, \gamma, \mu)$. We characterized the behavior of the system by drawing stability diagrams in the $(K,\gamma)$ plane, for differing values of $\mu$. 

For the broad influence function $P(\theta) = 1 + \cos \theta$, we found four qualitatively distinct stability diagrams. At all parameter values, the activity $\sigma(t)$ was ultimately stationary, although regions of bistability were identified. 

For the narrow influence function $P(\theta) = \delta(\theta)$, the bifurcation diagrams were similar to those for the broad pulse function, with one salient difference: the curve of Hopf bifurcations occurring for $K<0$ was now supercritical, giving rise to a stable limit cycle and the possibility of persistent oscillatory activity. 

This qualitative difference between the broad and narrow pulse regimes begs the question: what is the critical width of $P(\theta)$ that separates these two extremes? Future work could answer this question by studying a family of influence functions: $P_n(\theta) = a_n (1 + \cos \theta)^n$, where $a_n$ is a normalization constant, as in Refs.~\cite{pazo1, pazo2}. These functions get narrower as $n$ gets bigger, and represent our broad and narrow pulses as limits when $n=1$ and $n\rightarrow \infty$ respectively. 

Another interesting parameter to vary is the location of the maximum of the pulse function $P(\theta)$. We assumed $P(\theta)$  reaches its maximum at $\theta = 0$, the same phase where the excitability term $b \cos \theta$ is maximal.  Could removing this coincidence lead to new behavior? Other idealizations in the model could  also be relaxed. For example, one could introduce delay, mixed positive and negative coupling, nontrivial connectivity, and so on.


\begin{acknowledgments} 
Research supported in part by NSF grants DMS-1513179 and CCF-1522054.
\end{acknowledgments}


\end{document}